\documentclass[aps,onecolumn,preprint]{revtex4}

\usepackage{graphicx}
\usepackage{dcolumn}
\usepackage{natbib}
\usepackage{algorithm}
\usepackage{algorithmic}
\usepackage{amssymb}

\begin{document}

\title{Delayed Slater determinant update algorithms for high
  efficiency quantum Monte Carlo }
\thanks{
Notice: This manuscript has been authored by UT-Battelle, LLC under
Contract No. DE-AC05-00OR22725 with the U.S. Department of Energy.
The United States Government retains and the publisher, by accepting
the article for publication, acknowledges that the United States
Government retains a non-exclusive, paid-up, irrevocable, world-wide
license to publish or reproduce the published form of this manuscript,
or allow others to do so, for United States Government purposes.
The Department of Energy will provide public access to these results
of federally sponsored research in accordance with the DOE Public
Access Plan (\url{http://energy.gov/downloads/doe-public-access-plan}).
}

\author{T. McDaniel}
\email{bmcdan16@utk.edu}
\affiliation{Department of Electrical Engineering and Computer Science, University of Tennessee, Knoxville, TN 37996}
 
\author{E. F. D'Azevedo}%
\affiliation{Computer Science and Mathematics Division, Oak Ridge National Laboratory, Oak Ridge, TN 37831}

\author{Y. W. Li}
\affiliation{National Center for Computational Sciences, Oak Ridge National Laboratory, Oak Ridge, TN 37831}

\author{K. Wong}
\affiliation{Joint Institute for Computational Sciences, University of Tennessee, Knoxville, TN 37996}

\author{P. R. C. Kent}
\email{kentpr@ornl.gov}
\affiliation{Center for Nanophase Materials Sciences, Oak Ridge National Laboratory, Oak Ridge, TN 37831}
\affiliation{Computational Science and Engineering Division, Oak Ridge National Laboratory, Oak Ridge, TN 37831}

\date{\today}

\begin{abstract}
  Within ab initio Quantum Monte Carlo simulations, the leading
  numerical cost for large systems is the computation of the values of
  the Slater determinants in the trial wavefunction.  Each Monte Carlo
  step requires finding the determinant of a dense matrix. This is
  most commonly iteratively evaluated using a rank-1 Sherman-Morrison
  updating scheme to avoid repeated explicit calculation of the
  inverse. The overall computational cost is therefore formally cubic
  in the number of electrons or matrix size. To improve the numerical
  efficiency of this procedure, we propose a novel multiple rank
  delayed update scheme. This strategy enables probability evaluation
  with application of accepted moves to the matrices delayed until
  after a predetermined number of moves, $K$. The accepted events are
  then applied to the matrices \textit{en bloc} with enhanced
  arithmetic intensity and computational efficiency via matrix-matrix
  operations instead of matrix-vector operations. This procedure does
  not change the underlying Monte Carlo sampling or its statistical
  efficiency. For calculations on large systems and algorithms such as
  diffusion Monte Carlo where the acceptance ratio is high, order of
  magnitude improvements in the update time can be obtained on both
  multi-core CPUs and GPUs.
\end{abstract}

\maketitle

\section{Introduction}
\label{sec:introduction}

Ab initio Quantum Monte Carlo (QMC) techniques encompass a large class
of related methods and algorithms for many-body electronic structure
calculations~\cite{CeperleyPRB1977}. The variational Monte Carlo (VMC) and
diffusion Monte Carlo (DMC)
algorithms in particular have found wide application where the
potential accuracy of a full many-body calculation is preferred over 
mean field methods~\cite{foulkes_quantum_2001,wagner_discovering_2016}. 
Improvements in computational power as well as
important technical improvements to the methods are enabling
calculations on more complex systems, recently including cerium metal~\cite{devaux_electronic_2015},
copper oxide based superconductors~\cite{foyevtsova_ab_2014,wagner_effect_2014}, and
numerous binary transition metal
oxides~\cite{mitra_many-body_2015,santana_cohesive_2016,
yu_towards_2015,schiller_phase_2015,luo_phase_2016,
trail_quantum_2017,benali_quantum_2016,yu_fixed-node_2017}.

Central to all QMC algorithms is the procedure for proposing and
accepting Monte Carlo moves.  Each Monte Carlo step generally requires
proposing a move of a single electron, evaluating the probability of
the move, and updating the system state if the move is accepted. For
ab initio QMC calculations, Slater-Jastrow trial wave functions are
most commonly utilized, consisting of one or more Slater determinants
multiplied by a Jastrow correlation factor. In molecular calculations
it is now common to use large multideterminant
wavefunctions~\cite{guareschi_ground_2013,caffarel_communication_2016,filippi_simple_2016}
for small molecules with relatively low
electron counts. Great improvement in the 
evaluation of multideterminant wavefunctions has been made recently by 
evaluating the changes with respect to only a single
reference determinant~\cite{NukalaJCP2009,clark_computing_2011,filippi_simple_2016}, as opposed to
individually updating all the determinants, thereby enabling the
routine use of large multideterminant expansions in QMC. However, today, solid
state calculations most commonly use a single determinant and much larger
electron counts than molecular calculations due to the requirements of periodic boundary
conditions and need to perform finite size scaling. Evaluating these larger
determinants can occupy a significant fraction of overall computational time,
e.g. 30-50\% with electron counts in the thousands, indicating
that improved algorithms are desirable, particularly as larger solid
state calculations are performed and heavier elements with more
valence electrons are studied.

A Slater determinant for a system of $N$ electrons can be
represented by:
\[
\Psi({{\bf r}}_{1}, {{\bf r}}_{2},
\cdots ,{{\bf r}}_{N}) = \frac{1}{\sqrt{N!}} 
\left|\begin{array}{cccc}
\phi_{1}({{\bf r}}_{1}) & \phi_{1}({{\bf r}}_{2}) & \cdots & \phi_{1}({{\bf r}}_{N})\\ 
\phi_{2}({{\bf r}}_{1}) & \phi_{2}({{\bf r}}_{2}) & \cdots & \phi_{2}({{\bf r}}_{N}) \\ 
\vdots
& \vdots & \ddots & \vdots \\ 
\phi_{N}({{\bf r}}_{1}) & \phi_{N}({{\bf r}}_{2}) & \cdots & \phi_{N}({{\bf r}}_{N})
\end{array}\right|,
\]
where $\phi_{1, 2 \cdots n}$
represent single-particle wave functions and 
${{\bf r}}_{1, 2 \cdots N}$ represent particle coordinates.
Evaluating the ratios of this determinant as single
particle moves are proposed is key to efficient and affordable QMC
simulations. Explicit recalculation of the determinant is an $O(N^3)$ operation
which is infeasible to perform at every Monte Carlo step and for large $N$.

If $A$ is the Slater matrix and $A_{new}$ is
the matrix after the proposed particle move for particle at 
position ${\bf r}_k$ to $\tilde{{\bf r}}_k$, the probability $p$ of
accepting this move  is related to the ratio
of determinants, $ p \propto  \det(A_{new})/\det(A)$,
where $A_{new} = A + u  {v}_k'$,  $v_k$ is $k$-th column
of identity matrix, and column vector $u$, where
the $i$-th entry
$u_i = \left[ \phi_i({\tilde{{\bf r}}_k}) - \phi_i({{\bf r}_k}) \right], 1 \le i \le N$ 
is the 
change in the $k$-th column of matrix $A$.
This probability $p$
can be found via the Matrix
Determinant Formula,  
\begin{equation}  
{\det(A + u{v}_k') = \det(1 + {v}_k'A^{-1}u)  \det(A)} 
~.
\label{eqn:determinant}
\end{equation} 
Dividing this
equation on both sides by $\det(A)$, the existing determinant, yields
the Slater change ratio on the right-hand side. 

Note that $A^{-1} u$ is needed in the computation of 
determinant in Equation (\ref{eqn:determinant}).
The standard technique used in QMC~\cite{CeperleyPRB1977,fahy_variational_1990} 
is to construct and maintain
an explicit copy of the matrix inverse $A^{-1}$.
If the proposed move is accepted, the matrix inverse
can be updated using a rank-1 update using the Sherman-Morrison formula~\cite{GolubBook1996},
\begin{equation}
{(A + u v')^{-1} = A^{-1}- A^{-1}u (1 + v'A^{-1}u)^{-1}  v'A^{-1}}
~.
\label{eqn:sherman_morrison}
\end{equation}

The above procedure works well, but the performance of the rank-1 updating is
usually limited by the performance of the matrix-vector operations and memory bandwidth. Therefore it does not run very
efficiently on current computer architectures. A more compute
intensive algorithm is desirable, particularly as these operations
grow in relative cost with increasing $N$.
%

In this paper we propose a novel algorithm for computing updates to
the Slater determinant inverses. The underlying Monte Carlo algorithm
remains unchanged, but the numerical efficiency is 
significantly higher than the traditional algorithm for large
matrices. The new algorithm works by delaying the updates until a
large block update is eventually performed, allowing the use of
matrix-matrix algebra. Although the new algorithm can be used for any
QMC algorithm, the efficiency is highest where the acceptance ratio is
high, such as in the DMC calculations. These are usually the most
computationally expensive to perform. 

Although use of rank-1 updates is standard in QMC, we note that
blocking of these operations in rank-k like schemes has been used to
obtain significant performance gains in other fields~\cite{alvarez_new_2008,he_parallel_2013}.


\section{Algorithm}
\label{sec:algorithms}

We begin by observing that in many QMC calculations, the probability $p$
for accepting a move  is high, where $p \ge 0.99$. This is the case in
projection algorithms such as DMC where a short time step is required
to accurately sample the Green's function.
Since the expected average length of a contiguous sequence of
accepted moves is $p/(1-p)$, it is likely that
at least $K$ (where $K=32$, or $K=64$) consecutive moves will mostly be
accepted. 

One approach then proposes a sequence of up to $K$ moves and
computes the determinant ratio probability at each step using
a generalization of the Matrix Determinant Formula (\ref{eqn:determinant})
\begin{equation}
{\det(A + UV') = \det(I_{k} + V'A^{-1}U) \det(A)}
~,
\label{eqn:determinant2}
\end{equation} 
where ${I_k}$ is the $k$ by $k$ identity matrix,
$U$ is $N$  by $k$, and $V$ is $N$ by $k$.

This algorithm requires that after a particle move
is proposed and accepted, the move's application to the Slater matrix
is \textit{delayed}. We advance instead to finding the ratio for
the next proposed move. A group of accepted changes is queued in this manner until a
preset limit of unapplied moves $K$ is reached. 
The sequence of accepted moves is then applied to the Slater matrix together as a rank-k
update (for $k$ accepted moves) using the
Sherman-Morrison-Woodbury formula 
\begin{equation} 
(A + UV')^{-1} = A^{-1} -   
   A^{-1}U  (I_k + V'A^{-1}U)^{-1}   V'A^{-1}
~.
\label{eqn:woodbury}
\end{equation}

The rank-k update operation can be performed very efficiently as
a matrix-matrix multiplication operation using linear algebra
libraries such as the Basic Linear Algebra Subroutine Library
(BLAS). These are significantly more computationally intensive and
less memory bandwidth limited than conventional rank-1 updates, which
are matrix-vector operations. The
number of moves to delay by, $K$, can be chosen to optimize performance on
a specific computational architecture and for a specific problem size.

Note that the  matrix ${I_{k} + V'A^{-1}U}$ is, at most, $K \times K$, and the
calculation of its determinant is much less burdensome than recomputing
${\det(A)}$.  Up to $K$ of these smaller determinants
are computed prior to each update to ${A}$; we term this process
\textit{look-ahead}. Since the ${I_{k} + V'A^{-1}U}$
matrices are relatively small (for low values of $K$), note that in our initial
testing the determinant
look-ahead work is performed on CPU, even for our GPU-accelerated
implementations. 

It is important to see  that since every move has a
distinct, individual acceptance probability, our delayed update approach
is entirely equivalent to the classical method of performing each particle move
one move at a time, and this algorithm should not be interpreted  as a
multiple-particle algorithm. The Monte Carlo algorithm is unchanged
at the simulation level. Only the numerical implementation is changed.

An important question relates to the handling of rejected moves. One
possibility would be to perform the matrix updates whenever a rejected
move is encountered. However, this would not allow for significant
delay for methods such as VMC where the acceptance ratio $p\approx
0.25-0.75$. As we will see numerically in Sec. \ref{sec:results}, matrix update delays of tens
to hundreds of moves are desirable. Instead we prefer to treat rejected moves similarly to an
accepted move, only substituting the unchanged (``unmoved''') matrix columns. When
the number of accepted moves has reached the preset limit $K$, then 
the matrix inverse is updated using the Sherman-Morrison-Woodbury
formula in Equation \ref{eqn:woodbury} and dense matrix-matrix operations. 
This variant has the advantage that even if the probability
of acceptance is not high, say $p = 0.5$, the rank-k operation
can still maintain high performance by using effective values
of $K=32$ or $K=64$. While the performance is therefore proportional
to $p$, as we will see in Sec.\ref{sec:results}, the performance
increase compared to rank-1 updates is great enough that the new
scheme can be applied even where the acceptance is low 
($p < 0.5$).

Although not explored here, we note that the above variant also simplifies
the use of batched update operations across multiple walkers 
since each operation will be performing exactly the same amount of
work in the rank-$K$ update to the matrix inverse. Batching of update operations
for small $N$ is desirable on GPUs due to their greater concurrency requirements
than conventional CPUs~\cite{esler_accelerating_2012}. 
We also note
that the new algorithm could be combined with that of
Ref.~\cite{clark_computing_2011} for the multideterminant case.

\section{Performance results}
\label{sec:results}

To test the new algorithm we developed a test application that
assumes 100\% acceptance probability. The performance is therefore
closely representative of a DMC calculation with a small time step,
and typical of production DMC calculations where the majority of
compute time for QMC investigations is spent.
Depending on the sophistication of the final implementation, we  expect
QMC simulations with lower acceptance probability to obtain
proportionately lower performance.

We developed a prototype CPU implementation and also a prototype GPU implementation where the
work is shared between CPU and GPU. The implementation used the
current QMCPACK code~\cite{qmcpackurl} as a reference, which we expect is
similar to other QMC codes.  Vendor
optimized BLAS was used for the CPU timings. Our GPU implementation
performs only the final and most computationally intensive matrix updates
on the GPU. These matrices are stored in the GPU memory. The CPU is used for other steps of the algorithm.  For
both CPU and GPU performance we therefore expect that a somewhat higher
performance could be achieved with additional optimization. For example, the look ahead
operations are not threaded, penalizing performance for large $K$. For
small problem sizes, performance might also be sensitive to memory layout.
Simultaneous updates of multiple matrices or ``batching'', as is
performed by the current QMCPACK GPU
implementation~\cite{esler_accelerating_2012}, was not explored. The
batching of operations reduces GPU kernel launch overhead and might also be explored on CPUs
for improved memory prefetching. 

Performance is measured in Monte Carlo column updates 
per second. Each measurement involved a complete sweep of updates
through the full matrix of size $N$, with a full accounting of the
look ahead cost in the delayed updating scheme. Numerical values of
the determinants were checked versus direct calculation after a full
update sweep. For every combination of matrix update delay and
matrix size, we performed five runs to reduce any scatter from system
interruptions. Limited scatter in timings was observed for the
smallest $N=256$ and $N=512$. Scatter for larger matrices was negligible.
Timings were obtained on an Intel Xeon workstation with NVIDIA K40
GPUs using the Intel 2017 compilers, Intel MKL BLAS implementation,
and NVIDIA CUDA 8.0.  Because our algorithm improves memory bandwidth
utilization, we expect similar performance trends to be obtained on all
current and upcoming architectures from ARM, AMD, IBM, Intel, and NVIDIA
(etc.). 
 
\begin{figure}
  \centering
  \includegraphics{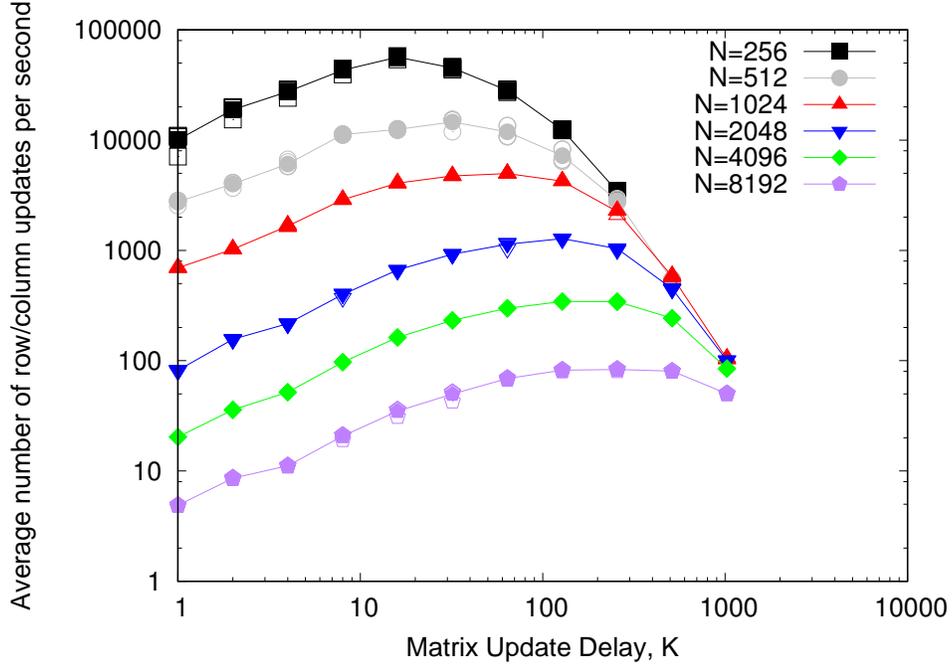}
  \caption{Single thread CPU performance of the delayed matrix
    updating scheme as a function of delay $K$, for matrix sizes $N$
    from 256 to 8192. The traditional Sherman-Morrison or rank-1
    algorithm corresponds to $K=1$. The lines pass through filled symbols indicating the average of five separate
    measurements at each delay and matrix size. Unfilled symbols give the individual timings, and show
    little variation. Speedups are found even for the smallest possible
    delay $K=2$ for all shown $N$.}
  \label{fig:cpu_single}
\end{figure}

\begin{figure}
  \centering
  \includegraphics{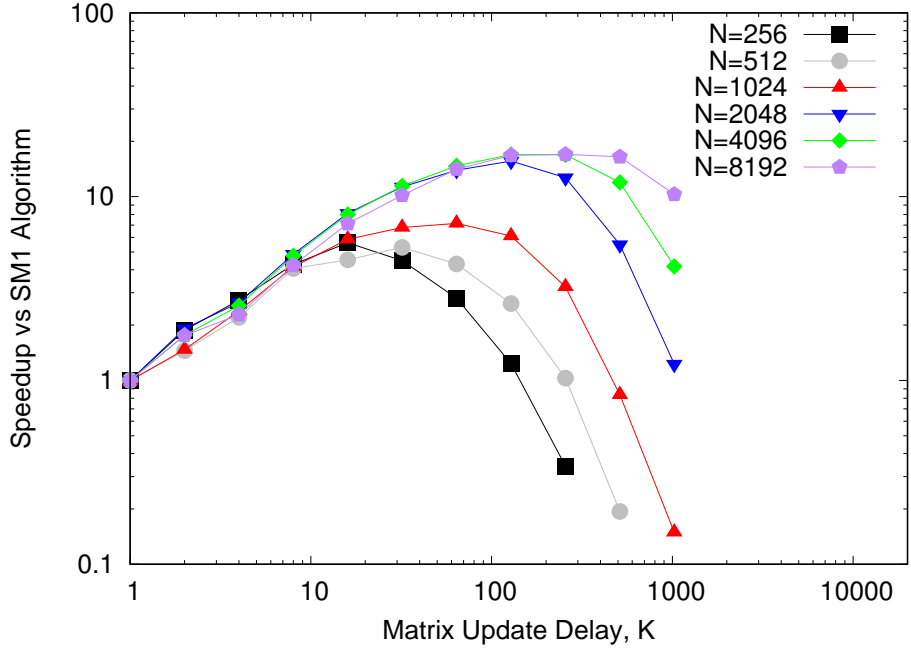}
  \caption{Relative speedup of single threaded CPU implementation of the delayed matrix
    updating scheme as a function of delay $K$, for matrix sizes $N$
    from $256$ to $8192$, relative to the Sherman-Morrison or rank-1
    algorithm. Highest speedups are $5.6 \times$ for $N=256$ with $K=16$,
    while for $N=8192$ and $K=256$ the measured speedup is $17.0 \times$
    over the conventional algorithm.}
  \label{fig:cpu_single_ratio}
\end{figure}

The absolute measured performance for different update delays and matrix sizes
is shown in Fig.~\ref{fig:cpu_single}. The proportionate
speedups relative to the conventional algorithm $K=1$ are shown in Fig.~\ref{fig:cpu_single_ratio}. 
The delayed update algorithm shows performance improvements for all
tested matrix sizes, with a generic shape to all of the performance
curves. For $N=256$ and $K=2$ we obtain a limited speedup of $1.8\times$,
increasing to an optimum $5.6\times$ for $K=16$. Above this the speedup is
reduced and is eventually lower than the conventional $K=1$
algorithm. This is because the naive explicit calculation of as many as $K$ determinants of
square matrices at most $K$ in size during the determinant look-ahead
process eventually becomes a significant computational burden. 
A refined implementation could delay this turnover to larger $K$. For
practical simulations the optimum $K$ could be chosen automatically
via an autotuner. 

With increased matrix size $N$ we observe in Fig.~\ref{fig:cpu_single_ratio} a general trend of greater
speedups relative to $K=1$ and a general trend to a higher delay
factor $K$ being optimal. Many of recently published QMC calculations have
$N \sim 512$ (for $\sim 1024$ total electrons, accounting for electron
spins), and an $N=2000$ calculation (4000 total electrons) was already
published in 2012~\cite{hood_diffusion_2012}.  For $N=1024$ we find an optimal
$K=64$ and speedup of $7.2\times$.

\begin{figure}
  \centering
  \includegraphics{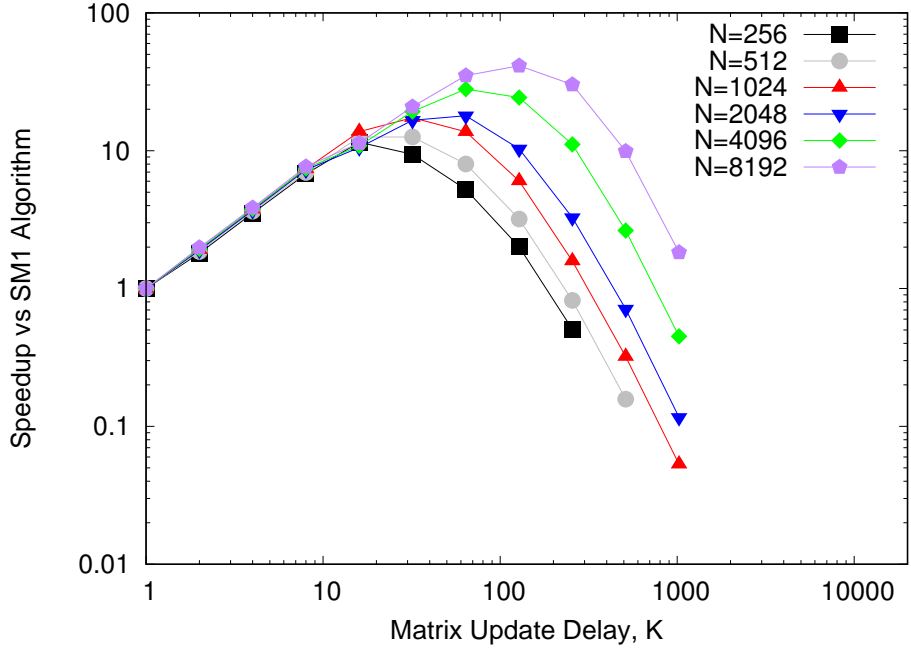}
  \caption{Relative speedup of GPU implementation of the delayed matrix
    updating scheme as a function of delay $K$, for matrix sizes $N$
    from $256$ to $8192$, relative to the Sherman-Morrison or rank-1
    algorithm. Highest speedups are $11.5 \times $ for $N=256$ using $K=16$,
    while for $N=8192$ and using $K=256$ the measured speedup is $41.3 \times $
    over the conventional algorithm.}
  \label{fig:gpu_single_ratio}
\end{figure}

The GPU implementation displayed very similar properties to the CPU
implementation in Fig.~\ref{fig:gpu_single_ratio}. Similar optimum $K$ are found as for the
CPU implementation. The peak speedups are
higher due the GPU numerical library's internal use of more threads in the matrix-matrix
operations, but the turnover to lower performance occurs more sharply
than the CPU implementation. This is in part due to the use of a
single threaded CPU-size implementation of the look-ahead process and
could be further optimized. 

\begin{figure}
  \centering
  \includegraphics{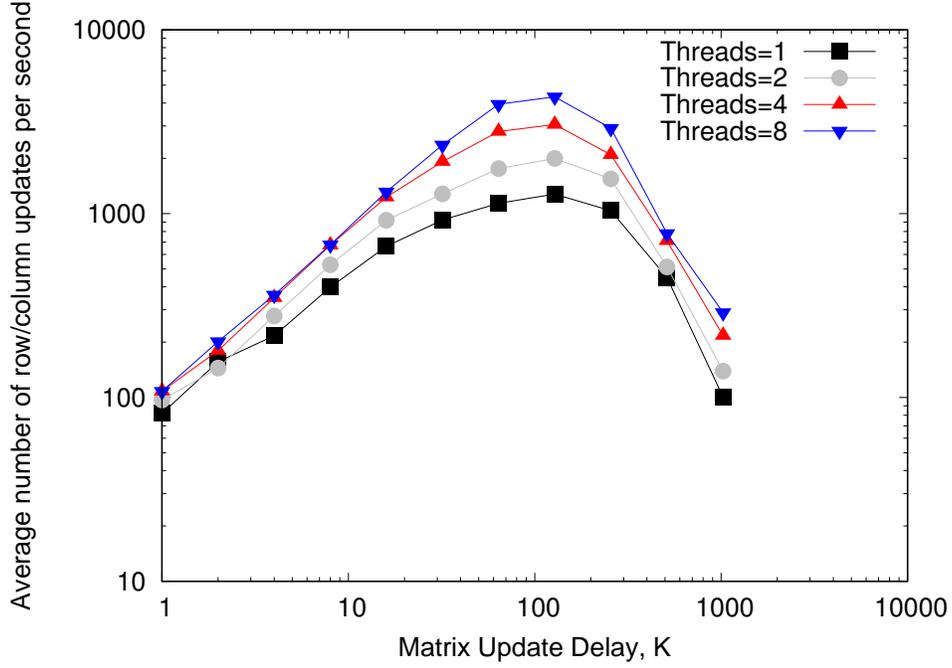}
  \caption{Multiple thread performance of CPU implementation of the
    delayed matrix updating scheme for
    matrix size $N=2048$.  The traditional Sherman-Morrison or rank-1
    algorithm corresponds to $K=1$. In this unoptimized
    implementation, only the final update operation is threaded. While
    threads have limited benefit for $K=1$, at optimal $K=128$ the
    delayed scheme results in increased threading efficiency and
    overall faster updates.}
  \label{fig:cpu_threads}
\end{figure}

With the aim of reducing the time per update further, we also
investigated use of multithreaded BLAS in the update operations,
relying solely on the vendor threaded libraries. As shown in
Fig.~\ref{fig:cpu_threads}, threading provides increased benefit
with large $K$. For $N=2048$, 8 threads improve the traditional $K=1$
Sherman-Morrison performance by $32\%$ relative to the single threaded
result. However, with optimum $K=128$ and the delayed scheme, 8 threads give a speedup of
$40.0\times$ relative to the threaded $K=1$ result, or $52.9\times$ over the
singly thread $K=1$ case. The delayed update algorithm therefore
significantly improves the utility of threads in reducing matrix update time
in QMC.

\section{Conclusions}
\label{sec:conclusions}


We have presented a simple and novel algorithm for improving the performance of updating the 
Slater determinants in QMC by using the determinant formula
to delay applying updates to the explicit matrix inverse. Despite the
increased cost of obtaining intermediate values, significant net speedups
are obtained over the traditional Sherman-Morrison algorithm. Order of magnitude speedups
are obtained with matrix sizes around 1000 on both CPU and GPU systems, with greater speedups obtained for larger matrices. These sizes
are similar to those of numerous recent solid state QMC
calculations. We encourage adoption in production QMC codes.

Raw data, scripts to recreate all figures in this manuscript, and
additional analyses for both CPU and GPU implementations are
available at \url{http://www.materialsdatafacility.org} and
\url{http://dx.doi.org/DATA_DOI_TO_BE_ADDED_ON_ACCEPTANCE}.

\section*{Acknowledgments}
TM was sponsored by the National Science Foundation
through Research Experience for Undergraduates (REU) 
award no. 1262937, with additional support from the National
Institute of Computational Sciences at University of Tennessee Knoxville. 
In addition, this work used allocations from the Extreme Science and 
Engineering Discovery Environment (XSEDE), which is supported by 
National Science Foundation grant number ACI-1053575.

EFD was supported by the U.S. Department of Energy Office of
Science, Advanced Scientific Computing Research (ASCR). 

YWL and computing resources were supported by the Oak Ridge 
Leadership Computing Facility, which is a DOE Office of Science User 
Facility supported under Contract DE-AC05-00OR22725.

PK was supported by the U.S. Department of Energy, Office of
Science, Basic Energy Sciences, Materials Sciences and Engineering
Division, as part of the Computational Materials Sciences Program and
Center for Predictive Simulation of Functional Materials.

\newpage

\begin{thebibliography}{26}
\expandafter\ifx\csname natexlab\endcsname\relax\def\natexlab#1{#1}\fi
\expandafter\ifx\csname bibnamefont\endcsname\relax
  \def\bibnamefont#1{#1}\fi
\expandafter\ifx\csname bibfnamefont\endcsname\relax
  \def\bibfnamefont#1{#1}\fi
\expandafter\ifx\csname citenamefont\endcsname\relax
  \def\citenamefont#1{#1}\fi
\expandafter\ifx\csname url\endcsname\relax
  \def\url#1{\texttt{#1}}\fi
\expandafter\ifx\csname urlprefix\endcsname\relax\def\urlprefix{URL }\fi
\providecommand{\bibinfo}[2]{#2}
\providecommand{\eprint}[2][]{\url{#2}}

\bibitem[{\citenamefont{Ceperley et~al.}(1977)\citenamefont{Ceperley, Chester,
  and Kalos}}]{CeperleyPRB1977}
\bibinfo{author}{\bibfnamefont{D.}~\bibnamefont{Ceperley}},
  \bibinfo{author}{\bibfnamefont{G.}~\bibnamefont{Chester}}, \bibnamefont{and}
  \bibinfo{author}{\bibfnamefont{M.}~\bibnamefont{Kalos}},
  \bibinfo{journal}{Physical Review B} \textbf{\bibinfo{volume}{16}},
  \bibinfo{pages}{3081} (\bibinfo{year}{1977}).

\bibitem[{\citenamefont{Foulkes et~al.}(2001)\citenamefont{Foulkes, Mitas,
  Needs, and Rajagopal}}]{foulkes_quantum_2001}
\bibinfo{author}{\bibfnamefont{W.~M.~C.} \bibnamefont{Foulkes}},
  \bibinfo{author}{\bibfnamefont{L.}~\bibnamefont{Mitas}},
  \bibinfo{author}{\bibfnamefont{R.~J.} \bibnamefont{Needs}}, \bibnamefont{and}
  \bibinfo{author}{\bibfnamefont{G.}~\bibnamefont{Rajagopal}},
  \bibinfo{journal}{Reviews of Modern Physics} \textbf{\bibinfo{volume}{73}},
  \bibinfo{pages}{33} (\bibinfo{year}{2001}).

\bibitem[{\citenamefont{Wagner and Ceperley}(2016)}]{wagner_discovering_2016}
\bibinfo{author}{\bibfnamefont{L.~K.} \bibnamefont{Wagner}} \bibnamefont{and}
  \bibinfo{author}{\bibfnamefont{D.~M.} \bibnamefont{Ceperley}},
  \bibinfo{journal}{Reports on Progress in Physics}
  \textbf{\bibinfo{volume}{79}}, \bibinfo{pages}{094501}
  (\bibinfo{year}{2016}).

\bibitem[{\citenamefont{Devaux et~al.}(2015)\citenamefont{Devaux, Casula,
  Decremps, and Sorella}}]{devaux_electronic_2015}
\bibinfo{author}{\bibfnamefont{N.}~\bibnamefont{Devaux}},
  \bibinfo{author}{\bibfnamefont{M.}~\bibnamefont{Casula}},
  \bibinfo{author}{\bibfnamefont{F.}~\bibnamefont{Decremps}}, \bibnamefont{and}
  \bibinfo{author}{\bibfnamefont{S.}~\bibnamefont{Sorella}},
  \bibinfo{journal}{Physical Review B} \textbf{\bibinfo{volume}{91}},
  \bibinfo{pages}{081101} (\bibinfo{year}{2015}).

\bibitem[{\citenamefont{Foyevtsova et~al.}(2014)\citenamefont{Foyevtsova,
  Krogel, Kim, Kent, Dagotto, and Reboredo}}]{foyevtsova_ab_2014}
\bibinfo{author}{\bibfnamefont{K.}~\bibnamefont{Foyevtsova}},
  \bibinfo{author}{\bibfnamefont{J.~T.} \bibnamefont{Krogel}},
  \bibinfo{author}{\bibfnamefont{J.}~\bibnamefont{Kim}},
  \bibinfo{author}{\bibfnamefont{P.}~\bibnamefont{Kent}},
  \bibinfo{author}{\bibfnamefont{E.}~\bibnamefont{Dagotto}}, \bibnamefont{and}
  \bibinfo{author}{\bibfnamefont{F.~A.} \bibnamefont{Reboredo}},
  \bibinfo{journal}{Physical Review X} \textbf{\bibinfo{volume}{4}},
  \bibinfo{pages}{031003} (\bibinfo{year}{2014}).

\bibitem[{\citenamefont{Wagner and Abbamonte}(2014)}]{wagner_effect_2014}
\bibinfo{author}{\bibfnamefont{L.~K.} \bibnamefont{Wagner}} \bibnamefont{and}
  \bibinfo{author}{\bibfnamefont{P.}~\bibnamefont{Abbamonte}},
  \bibinfo{journal}{Physical Review B} \textbf{\bibinfo{volume}{90}},
  \bibinfo{pages}{125129} (\bibinfo{year}{2014}).

\bibitem[{\citenamefont{Mitra et~al.}(2015)\citenamefont{Mitra, Krogel,
  Santana, and Reboredo}}]{mitra_many-body_2015}
\bibinfo{author}{\bibfnamefont{C.}~\bibnamefont{Mitra}},
  \bibinfo{author}{\bibfnamefont{J.~T.} \bibnamefont{Krogel}},
  \bibinfo{author}{\bibfnamefont{J.~A.} \bibnamefont{Santana}},
  \bibnamefont{and} \bibinfo{author}{\bibfnamefont{F.~A.}
  \bibnamefont{Reboredo}}, \bibinfo{journal}{Journal of Chemical Physics}
  \textbf{\bibinfo{volume}{143}}, \bibinfo{pages}{164710}
  (\bibinfo{year}{2015}).

\bibitem[{\citenamefont{Santana et~al.}(2016)\citenamefont{Santana, Krogel,
  Kent, and Reboredo}}]{santana_cohesive_2016}
\bibinfo{author}{\bibfnamefont{J.~A.} \bibnamefont{Santana}},
  \bibinfo{author}{\bibfnamefont{J.~T.} \bibnamefont{Krogel}},
  \bibinfo{author}{\bibfnamefont{P.~R.~C.} \bibnamefont{Kent}},
  \bibnamefont{and} \bibinfo{author}{\bibfnamefont{F.~A.}
  \bibnamefont{Reboredo}}, \bibinfo{journal}{Journal of Chemical Physics}
  \textbf{\bibinfo{volume}{144}}, \bibinfo{pages}{174707}
  (\bibinfo{year}{2016}).

\bibitem[{\citenamefont{Yu et~al.}(2015)\citenamefont{Yu, Wagner, and
  Ertekin}}]{yu_towards_2015}
\bibinfo{author}{\bibfnamefont{J.}~\bibnamefont{Yu}},
  \bibinfo{author}{\bibfnamefont{L.~K.} \bibnamefont{Wagner}},
  \bibnamefont{and} \bibinfo{author}{\bibfnamefont{E.}~\bibnamefont{Ertekin}},
  \bibinfo{journal}{Journal of Chemical Physics}
  \textbf{\bibinfo{volume}{143}}, \bibinfo{pages}{224707}
  (\bibinfo{year}{2015}).

\bibitem[{\citenamefont{Schiller et~al.}(2015)\citenamefont{Schiller, Wagner,
  and Ertekin}}]{schiller_phase_2015}
\bibinfo{author}{\bibfnamefont{J.~A.} \bibnamefont{Schiller}},
  \bibinfo{author}{\bibfnamefont{L.~K.} \bibnamefont{Wagner}},
  \bibnamefont{and} \bibinfo{author}{\bibfnamefont{E.}~\bibnamefont{Ertekin}},
  \bibinfo{journal}{Physical Review B} \textbf{\bibinfo{volume}{92}},
  \bibinfo{pages}{235209} (\bibinfo{year}{2015}).

\bibitem[{\citenamefont{Luo et~al.}(2016)\citenamefont{Luo, Benali,
  Shulenburger, Krogel, Heinonen, and Kent}}]{luo_phase_2016}
\bibinfo{author}{\bibfnamefont{Y.}~\bibnamefont{Luo}},
  \bibinfo{author}{\bibfnamefont{A.}~\bibnamefont{Benali}},
  \bibinfo{author}{\bibfnamefont{L.}~\bibnamefont{Shulenburger}},
  \bibinfo{author}{\bibfnamefont{J.~T.} \bibnamefont{Krogel}},
  \bibinfo{author}{\bibfnamefont{O.}~\bibnamefont{Heinonen}}, \bibnamefont{and}
  \bibinfo{author}{\bibfnamefont{P.~R.~C.} \bibnamefont{Kent}},
  \bibinfo{journal}{New Journal of Physics} \textbf{\bibinfo{volume}{18}},
  \bibinfo{pages}{113049} (\bibinfo{year}{2016}).

\bibitem[{\citenamefont{Trail et~al.}(2017)\citenamefont{Trail, Monserrat,
  López~Ríos, Maezono, and Needs}}]{trail_quantum_2017}
\bibinfo{author}{\bibfnamefont{J.}~\bibnamefont{Trail}},
  \bibinfo{author}{\bibfnamefont{B.}~\bibnamefont{Monserrat}},
  \bibinfo{author}{\bibfnamefont{P.}~\bibnamefont{López~Ríos}},
  \bibinfo{author}{\bibfnamefont{R.}~\bibnamefont{Maezono}}, \bibnamefont{and}
  \bibinfo{author}{\bibfnamefont{R.~J.} \bibnamefont{Needs}},
  \bibinfo{journal}{Physical Review B} \textbf{\bibinfo{volume}{95}},
  \bibinfo{pages}{121108} (\bibinfo{year}{2017}).

\bibitem[{\citenamefont{Benali et~al.}(2016)\citenamefont{Benali, Shulenburger,
  Krogel, Zhong, Kent, and Heinonen}}]{benali_quantum_2016}
\bibinfo{author}{\bibfnamefont{A.}~\bibnamefont{Benali}},
  \bibinfo{author}{\bibfnamefont{L.}~\bibnamefont{Shulenburger}},
  \bibinfo{author}{\bibfnamefont{J.~T.} \bibnamefont{Krogel}},
  \bibinfo{author}{\bibfnamefont{X.}~\bibnamefont{Zhong}},
  \bibinfo{author}{\bibfnamefont{P.~R.~C.} \bibnamefont{Kent}},
  \bibnamefont{and} \bibinfo{author}{\bibfnamefont{O.}~\bibnamefont{Heinonen}},
  \bibinfo{journal}{Phys. Chem. Chem. Phys.} \textbf{\bibinfo{volume}{18}},
  \bibinfo{pages}{18323} (\bibinfo{year}{2016}).

\bibitem[{\citenamefont{Yu et~al.}(2017)\citenamefont{Yu, Wagner, and
  Ertekin}}]{yu_fixed-node_2017}
\bibinfo{author}{\bibfnamefont{J.}~\bibnamefont{Yu}},
  \bibinfo{author}{\bibfnamefont{L.~K.} \bibnamefont{Wagner}},
  \bibnamefont{and} \bibinfo{author}{\bibfnamefont{E.}~\bibnamefont{Ertekin}},
  \bibinfo{journal}{Physical Review B} \textbf{\bibinfo{volume}{95}},
  \bibinfo{pages}{075209} (\bibinfo{year}{2017}).

\bibitem[{\citenamefont{Guareschi and Filippi}(2013)}]{guareschi_ground_2013}
\bibinfo{author}{\bibfnamefont{R.}~\bibnamefont{Guareschi}} \bibnamefont{and}
  \bibinfo{author}{\bibfnamefont{C.}~\bibnamefont{Filippi}},
  \bibinfo{journal}{Journal of Chemical Theory and Computation}
  \textbf{\bibinfo{volume}{9}}, \bibinfo{pages}{5513} (\bibinfo{year}{2013}).

\bibitem[{\citenamefont{Caffarel et~al.}(2016)\citenamefont{Caffarel,
  Applencourt, Giner, and Scemama}}]{caffarel_communication_2016}
\bibinfo{author}{\bibfnamefont{M.}~\bibnamefont{Caffarel}},
  \bibinfo{author}{\bibfnamefont{T.}~\bibnamefont{Applencourt}},
  \bibinfo{author}{\bibfnamefont{E.}~\bibnamefont{Giner}}, \bibnamefont{and}
  \bibinfo{author}{\bibfnamefont{A.}~\bibnamefont{Scemama}},
  \bibinfo{journal}{Journal of Chemical Physics}
  \textbf{\bibinfo{volume}{144}}, \bibinfo{pages}{151103}
  (\bibinfo{year}{2016}).

\bibitem[{\citenamefont{Filippi et~al.}(2016)\citenamefont{Filippi, Assaraf,
  and Moroni}}]{filippi_simple_2016}
\bibinfo{author}{\bibfnamefont{C.}~\bibnamefont{Filippi}},
  \bibinfo{author}{\bibfnamefont{R.}~\bibnamefont{Assaraf}}, \bibnamefont{and}
  \bibinfo{author}{\bibfnamefont{S.}~\bibnamefont{Moroni}},
  \bibinfo{journal}{Journal of Chemical Physics}
  \textbf{\bibinfo{volume}{144}}, \bibinfo{pages}{194105}
  (\bibinfo{year}{2016}).

\bibitem[{\citenamefont{Nukala and Kent}(2009)}]{NukalaJCP2009}
\bibinfo{author}{\bibfnamefont{P.~K.} \bibnamefont{Nukala}} \bibnamefont{and}
  \bibinfo{author}{\bibfnamefont{P.}~\bibnamefont{Kent}},
  \bibinfo{journal}{Journal of Chemical Physics}
  \textbf{\bibinfo{volume}{130}}, \bibinfo{pages}{204105}
  (\bibinfo{year}{2009}).

\bibitem[{\citenamefont{Clark et~al.}(2011)\citenamefont{Clark, Morales,
  {McMinis}, Kim, and Scuseria}}]{clark_computing_2011}
\bibinfo{author}{\bibfnamefont{B.~K.} \bibnamefont{Clark}},
  \bibinfo{author}{\bibfnamefont{M.~A.} \bibnamefont{Morales}},
  \bibinfo{author}{\bibfnamefont{J.}~\bibnamefont{{McMinis}}},
  \bibinfo{author}{\bibfnamefont{J.}~\bibnamefont{Kim}}, \bibnamefont{and}
  \bibinfo{author}{\bibfnamefont{G.~E.} \bibnamefont{Scuseria}},
  \bibinfo{journal}{Journal of Chemical Physics}
  \textbf{\bibinfo{volume}{135}}, \bibinfo{pages}{244105}
  (\bibinfo{year}{2011}).

\bibitem[{\citenamefont{Fahy et~al.}(1990)\citenamefont{Fahy, Wang, and
  Louie}}]{fahy_variational_1990}
\bibinfo{author}{\bibfnamefont{S.}~\bibnamefont{Fahy}},
  \bibinfo{author}{\bibfnamefont{X.~W.} \bibnamefont{Wang}}, \bibnamefont{and}
  \bibinfo{author}{\bibfnamefont{S.~G.} \bibnamefont{Louie}},
  \bibinfo{journal}{Physical Review B} \textbf{\bibinfo{volume}{42}},
  \bibinfo{pages}{3503} (\bibinfo{year}{1990}).

\bibitem[{\citenamefont{Golub and Van~Loan}(1996)}]{GolubBook1996}
\bibinfo{author}{\bibfnamefont{G.~H.} \bibnamefont{Golub}} \bibnamefont{and}
  \bibinfo{author}{\bibfnamefont{C.~F.} \bibnamefont{Van~Loan}},
  \emph{\bibinfo{title}{Matrix Computations}} (\bibinfo{publisher}{Johns
  Hopkins Univ. Press}, \bibinfo{address}{Baltimore, M.A.},
  \bibinfo{year}{1996}).

\bibitem[{\citenamefont{Alvarez et~al.}(2008)\citenamefont{Alvarez, Summers,
  Maxwell, Eisenbach, Meredith, Larkin, Levesque, Maier, Kent, D'Azevedo
  et~al.}}]{alvarez_new_2008}
\bibinfo{author}{\bibfnamefont{G.}~\bibnamefont{Alvarez}},
  \bibinfo{author}{\bibfnamefont{M.~S.} \bibnamefont{Summers}},
  \bibinfo{author}{\bibfnamefont{D.~E.} \bibnamefont{Maxwell}},
  \bibinfo{author}{\bibfnamefont{M.}~\bibnamefont{Eisenbach}},
  \bibinfo{author}{\bibfnamefont{J.~S.} \bibnamefont{Meredith}},
  \bibinfo{author}{\bibfnamefont{J.~M.} \bibnamefont{Larkin}},
  \bibinfo{author}{\bibfnamefont{J.}~\bibnamefont{Levesque}},
  \bibinfo{author}{\bibfnamefont{T.~A.} \bibnamefont{Maier}},
  \bibinfo{author}{\bibfnamefont{P.~R.~C.} \bibnamefont{Kent}},
  \bibinfo{author}{\bibfnamefont{E.~F.} \bibnamefont{D'Azevedo}},
  \bibnamefont{et~al.}, in \emph{\bibinfo{booktitle}{Proceedings of the 2008
  {ACM}/{IEEE} conference on Supercomputing}} (\bibinfo{publisher}{{IEEE}
  Press}, \bibinfo{year}{2008}), pp. \bibinfo{pages}{1--10}.

\bibitem[{\citenamefont{He et~al.}(2013)\citenamefont{He, Holm, and
  Neytcheva}}]{he_parallel_2013}
\bibinfo{author}{\bibfnamefont{X.}~\bibnamefont{He}},
  \bibinfo{author}{\bibfnamefont{M.}~\bibnamefont{Holm}}, \bibnamefont{and}
  \bibinfo{author}{\bibfnamefont{M.}~\bibnamefont{Neytcheva}}, in
  \emph{\bibinfo{booktitle}{Applied Parallel and Scientific Computing}}, edited
  by \bibinfo{editor}{\bibfnamefont{P.}~\bibnamefont{Manninen}}
  \bibnamefont{and} \bibinfo{editor}{\bibfnamefont{P.}~\bibnamefont{Öster}}
  (\bibinfo{publisher}{Springer Berlin Heidelberg}, \bibinfo{year}{2013}), no.
  \bibinfo{number}{7782} in \bibinfo{series}{Lecture Notes in Computer
  Science}, pp. \bibinfo{pages}{206--219}.

\bibitem[{\citenamefont{Esler et~al.}(2012)\citenamefont{Esler, Kim, Ceperley,
  and Shulenburger}}]{esler_accelerating_2012}
\bibinfo{author}{\bibfnamefont{K.}~\bibnamefont{Esler}},
  \bibinfo{author}{\bibfnamefont{J.}~\bibnamefont{Kim}},
  \bibinfo{author}{\bibfnamefont{D.}~\bibnamefont{Ceperley}}, \bibnamefont{and}
  \bibinfo{author}{\bibfnamefont{L.}~\bibnamefont{Shulenburger}},
  \textbf{\bibinfo{volume}{14}}, \bibinfo{pages}{40 } (\bibinfo{year}{2012}).

\bibitem[{qmc()}]{qmcpackurl}
\bibinfo{note}{\url{http://www.qmcpack.org}}.

\bibitem[{\citenamefont{Hood et~al.}(2012)\citenamefont{Hood, Kent, and
  Reboredo}}]{hood_diffusion_2012}
\bibinfo{author}{\bibfnamefont{R.~Q.} \bibnamefont{Hood}},
  \bibinfo{author}{\bibfnamefont{P.~R.~C.} \bibnamefont{Kent}},
  \bibnamefont{and} \bibinfo{author}{\bibfnamefont{F.~A.}
  \bibnamefont{Reboredo}}, \bibinfo{journal}{Physical Review B}
  \textbf{\bibinfo{volume}{85}}, \bibinfo{pages}{134109}
  (\bibinfo{year}{2012}).

\end{thebibliography}

\newpage

\end{document}